\documentclass{egpubl}
\usepackage{eurovis2022}

\EuroVisPoster % for EuroVis additional material 
\usepackage[T1]{fontenc}
\usepackage{dfadobe}  

\usepackage{cite}  % comment out for biblatex with backend=biber
\BibtexOrBiblatex
\electronicVersion
\PrintedOrElectronic
\ifpdf \usepackage[pdftex]{graphicx} \pdfcompresslevel=9
\else \usepackage[dvips]{graphicx} \fi

\usepackage{egweblnk}
\usepackage{flushend}
\usepackage[dvipsnames]{xcolor}

\newlength{\maxlen}
\usepackage{tikzsymbols}
	
\newlength{\mytextsize}

\newcommand{\bpstart}[1]{\vspace{1mm} \noindent{\textbf{#1}}}
\newcommand{\inlinevis}[3]{\raisebox{#1}[0pt][0pt]{\includegraphics[height=#2]{#3}}}

\usepackage{svg}
\usepackage{amsmath}
\usepackage{wrapfig}
\usepackage{microtype}
\title[Situated Visualization in Motion for Video games]%
      {Situated Visualization in Motion for Video games}

\author[Federica Bucchieri, Lijie Yao, \& Petra Isenberg]
{\parbox{\textwidth}{
    \centering 
        Federica Bucchieri$^{1}$\orcid{0009-0009-6398-0660},
        Lijie Yao$^{1}$\orcid{0000-0002-4208-5140},
        and Petra Isenberg$^{1}$\orcid{0000-0002-2948-6417} 
        }
        \\
{\parbox{\textwidth}{
    \centering 
        $^1$Universit{\'e} Paris-Saclay, CNRS, Inria, LISN, Orsay, France
      }
}
}
\begin{document}

% uncomment for using teaser
% \teaser{
%  \includegraphics[width=\linewidth]{eg_new}
%  \centering
%   \caption{New EG Logo}
% \label{fig:teaser}
%}

\maketitle
%-------------------------------------------------------------------------
\begin{abstract}
   %What
   We contribute a systematic review of situated visualizations in motion in the context of video games. 
   %Why
   Video games produce rich dynamic datasets during gameplay that are often visualized to help players succeed in a game. Often these visualizations are moving either because they are attached to moving game elements or due to camera changes. We want to understand to what extent this motion and contextual game factors impact how players can read these visualizations.  
   %How
   In order to ground our work, we surveyed 160 visualizations in motion and their embeddings in the game world. Here, we report on our analysis and categorization of these visualizations. 
%-------------------------------------------------------------------------
%  ACM CCS 1998
% (see https://www.acm.org/publications/computing-classification-system/1998)
% \begin{classification} % according to https://www.acm.org/publications/computing-classification-system/1998
% \CCScat{Computer Graphics}{I.3.3}{Picture/Image Generation}{Line and curve generation}
% \end{classification}
%-------------------------------------------------------------------------
%  ACM CCS 2012
%   (see https://www.acm.org/publications/class-2012)
%The tool at \url{http://dl.acm.org/ccs.cfm} can be used to generate
% CCS codes.
%Example:


\end{abstract}  
%-------------------------------------------------------------------------

%------ REAL VIDEO GAMES IMAGES AND CAPTION
%\begin{figure*}[tbp]
 % \centering
  % the following command controls the width of the embedded PS file
  % (relative to the width of the current column)
  %\includegraphics[width=.328\linewidth]{egPublStyle-EuroVis_full-short-stars-posters-2022/Fig 1 - (a) Real game.png}
  %\hfill
  %\includegraphics[width=.328\linewidth]{egPublStyle-EuroVis_full-short-stars-posters-2022/Fig 1 - (b) Real game.png}
%\hfill
 % \includegraphics[width=.328\linewidth]{egPublStyle-EuroVis_full-short-stars-posters-2022/Fig 1 - (c) Real game.png} 
%\caption{\label{fig:ex1}%
           %Different embedding locations of visualization in motion. Left: Stamina bar placed below the characters in NBA 2K21 \cite{NBA2K21}. Middle: Circular health bar placed over an enemy in The Falconeer \cite{Falconeer}. Right: Ammunition count integrated in the weapon's (character) design in Halo Infinite \cite{Halo}.
 %          }
%\end{figure*}

%-------------------------------------------------------------------------

\section{Introduction}
With the prevalence of video games and the growing games market, many people are used to following moving game elements on the screen. Gameplay produces rich and changing data, some of which is shown to gamers in the form of visualizations that are also often moving.  These visualizations are designed to aid players succeed in the game and make important decisions about how to act in the game. Examples include health bars, navigation aids, resource inventories, or team membership. As such, visualizations play an important role in how effective a player can be but they also pose a number of interesting design challenges. Visualizations need to be read at a glance while the player is focused on a primary task such as fulfilling a game mission. They also often need to be small, match the aesthetics of the game, or be closely embedded next to game elements. In order to survey the current practice of visualizations in  motion in video games, we reviewed 50 games from 17 different genres. In these games we found 160 examples of visualizations in motion that we categorize according to different dimensions. Our future research goal is to design and embed visualizations in motion in the context of games and to explore the impact of contextual factors in video games on visualizations in motion.

%Games have been the topic of some research in the visualization community \cite{Zamitto:2008:VisTechInVideoGames, Bowman:2012:TowardVisforGames, Peacocke:2018:EmpiricalComparisonFPS}. Our work relates most closely to the space of \emph{visualization in motion} as proposed by Yao et al.\cite{Yao:2020:SituatedVisinMotion} and situated visualizations \cite{Willett:2017:EDR}.

%-------------------------------------------------------------------------
\vspace{-10pt}
\section{Related Work}
Yao et al. \cite{Yao:2020:SituatedVisinMotion} proposed a first design space for \emph{visualization in motion} that included the motion relationship between viewer and visualization.  When playing video games this motion relationship involves a stationary player sitting in front of a monitor, seeing visualizations moving on the screen. Visualizations are \emph{situated} in the game context and as such our work related to the research area of situated visualization. White \cite{White:2009:IPTforSituatedVis} studied different presentation and interaction techniques, investigating a theoretical framework and best practises for situated visualizations. More recently Willett et al. \cite{Willett:2017:EDR} formalized the difference between situated and embedded visualization, analyzing the relationship among data and the referents to which the data refers. Our work focuses on embedded visualizations that have a close connection between data and referent; for example where data about a game character is displayed directly next to the character's representation.  Our work is of course, also related to past work on visualization for video games. 
Zammitto \cite{Zamitto:2008:VisTechInVideoGames} was the first one to tackle the topic and analyze the principles of visualizations used in games. Her analysis focused on determining how video games provide the player with important visual information like the use of silhouettes, mini-maps, HUDs, Fog of war etc. Some theories and design spaces have been proposed, based on a systematic review of existing games that incorporate some sort of visualization. Bowman's \cite{Bowman:2012:TowardVisforGames} framework comprises five categories to classify any video games visualization: primary purpose, target audience, temporal usage, visual complexity and immersion/integration.  Peacocke et al. \cite{Peacocke:2018:EmpiricalComparisonFPS} studied players' performances with different types of displays to find the best way to represent different types of data. Their work shows that there is no universally best display type for every information. Different representations worked well for specific types of information (e.g. players performed better with information about ammo displayed with spatial displays). Our research deviates from this past work in that we concentrate on the motion of visualizations related to contextual factors of the game itself.

%-------------------------------------------------------------------------
\vspace{-10pt}
\section{A Review of Visualizations in Motion in Video Games}
To ground our work in existing practices of moving visualizations in video games we conducted a systematic review. To cover a diverse selection of video games we made use of a commercial ranking website called Metacritic \cite{Metacritic}. 
Metacritic assigns a unique Metascore to each video game, which is a weighted average of the scores of the world's most respected reviewers from the gaming field.
Metacritic categorizes games according to 18 different game genres. For each genre we selected the top 3 games from 2011 to 2022 sorted by the Metacritic relevance score. We considered all gaming platforms such as PC, Xbox and PlayStation etc. We excluded puzzle games because they did not contain moving visualizations. Moreover, out of the first 50 wrestling games only two provided relevant visualizations. In total, we reviewed 50 games from 17 genres. For each game, one author of this poster watched game-plays on YouTube for approximately 5 to 15 minutes and video-recorded relevant parts of the videos where the game showed visualizations in motion. % I tried to fix it  ....

%-------------------------------------------------------------------------
\vspace{-10pt}
\section{Categorization of Current Video Game Visualizations}
\begin{figure}[htb]
  \centering
  \includegraphics[width=.325\linewidth]{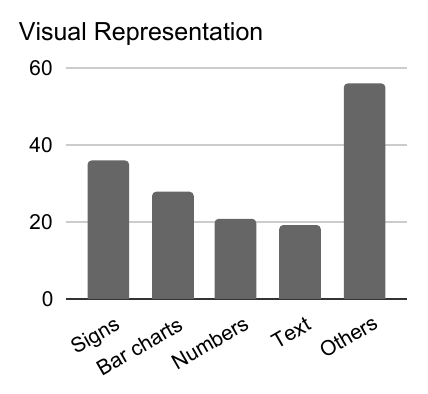}
   \includegraphics[width=.325\linewidth]{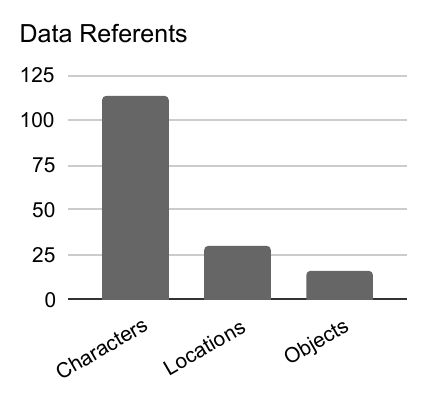}
   \includegraphics[width=.325\linewidth]{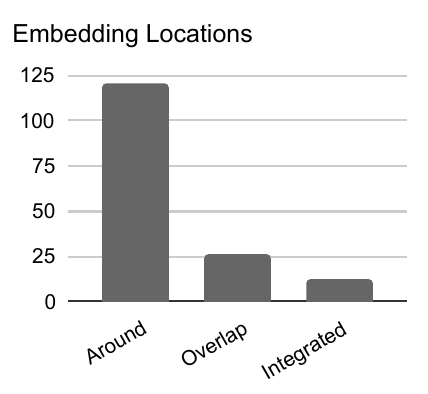}
  \caption{\label{fig:ex3}Counts for the number of examples including factors of visual representation, data referents, and embedding locations.}
\end{figure}

In total, we collect 160 examples of visualization in motion. We categorized these examples according to multiple dimensions related to situated visualization and motion characteristics:

\bpstart{Visualization Representation:} describes how data was represented. Signs \inlinevis{-2pt}{1.2em}{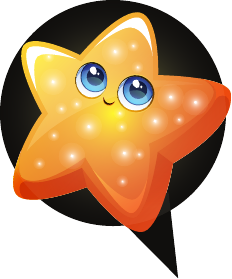} were the most prevalent representation (36/160), followed by bar charts %\inlinevis{-2pt}{1.2em}{radialbar} 
\inlinevis{0pt}{0.5em}{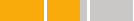} (28/160), labels with numbers (21/160) and labels with texts (19/160). The remaining 56/160 visualizations included silhouettes, pictographs and more (see \autoref{fig:ex3}: Left).

\begin{figure}[htb]
  \centering
  % the following command controls the width of the embedded PS file
  % (relative to the width of the current column)
  \includegraphics[width=.325\linewidth]{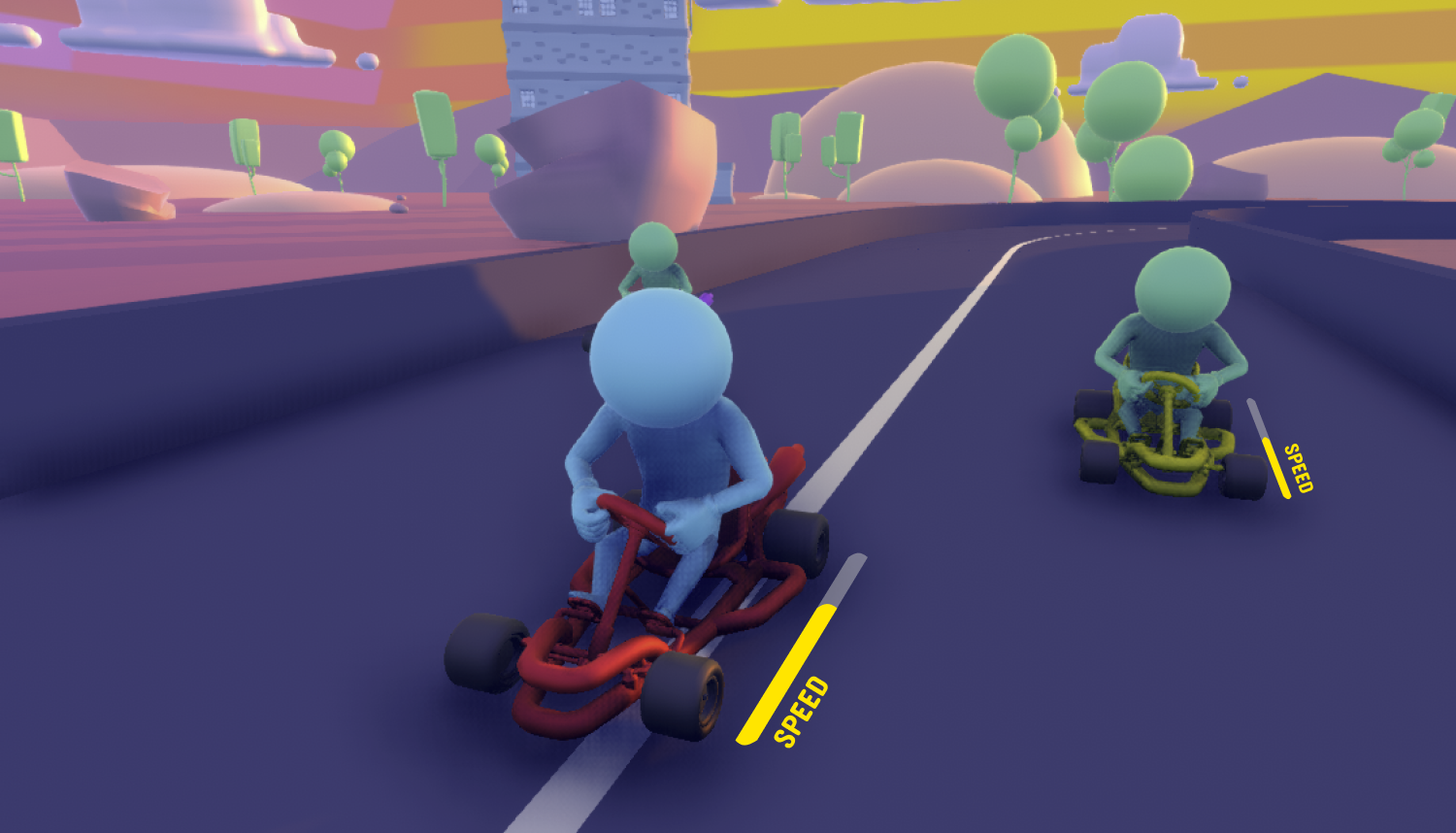}
  \hfill
  \includegraphics[width=.325\linewidth]{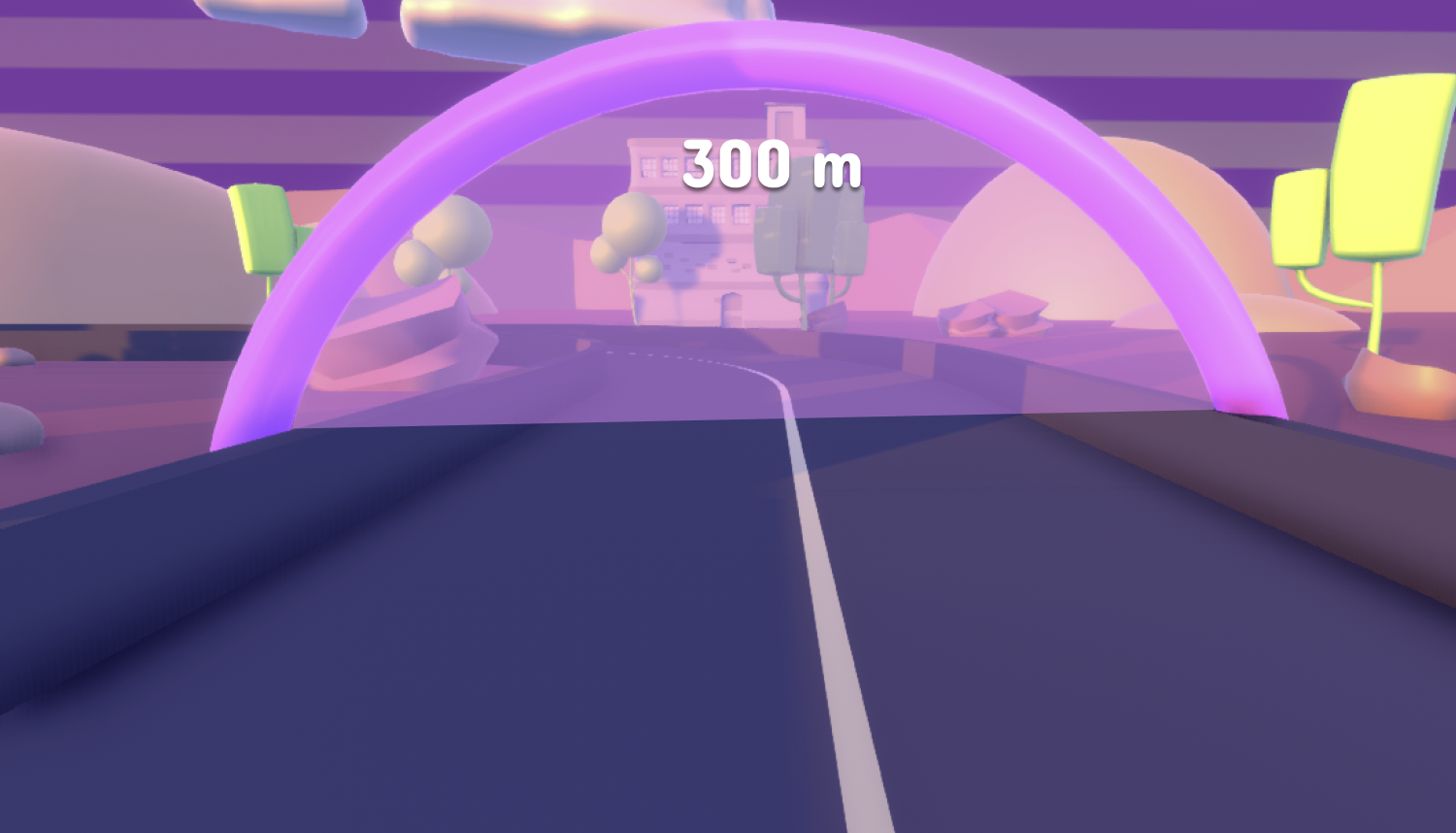}
\hfill
  \includegraphics[width=.325\linewidth]{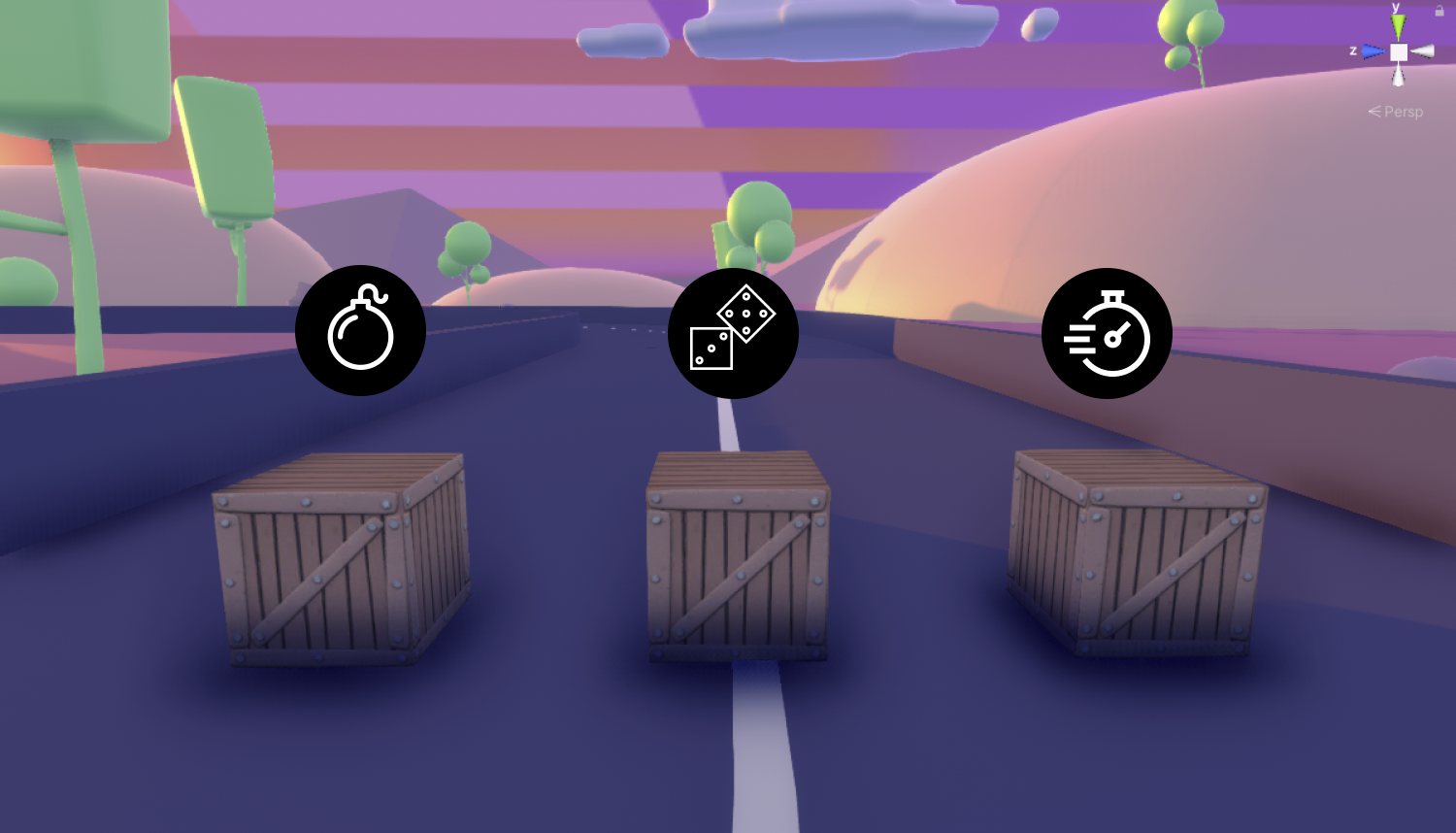}
  \caption{\label{fig:ex1}%
            Visualizations referring to different data referents in a racing game. Left: characters' speed bars. Middle: checkpoint indicator with text label showing distance. Right: signs for object type.
          }

\end{figure}

\vspace{-15pt}
\bpstart{Data referents:} are the entities that the data refers to. Most referents are game characters (see \autoref{fig:ex1}: Left), locations (see \autoref{fig:ex1}: Middle), or an object in the game (see \autoref{fig:ex1}: Right). In our sample collection, 114/160 data referents were game characters, 30/160 were locations, and 16/160 were objects (see \autoref{fig:ex3}: Middle).

\begin{figure}[h]
  \centering
  % the following command controls the width of the embedded PS file
  % (relative to the width of the current column)
  \includegraphics[width=.325\linewidth]{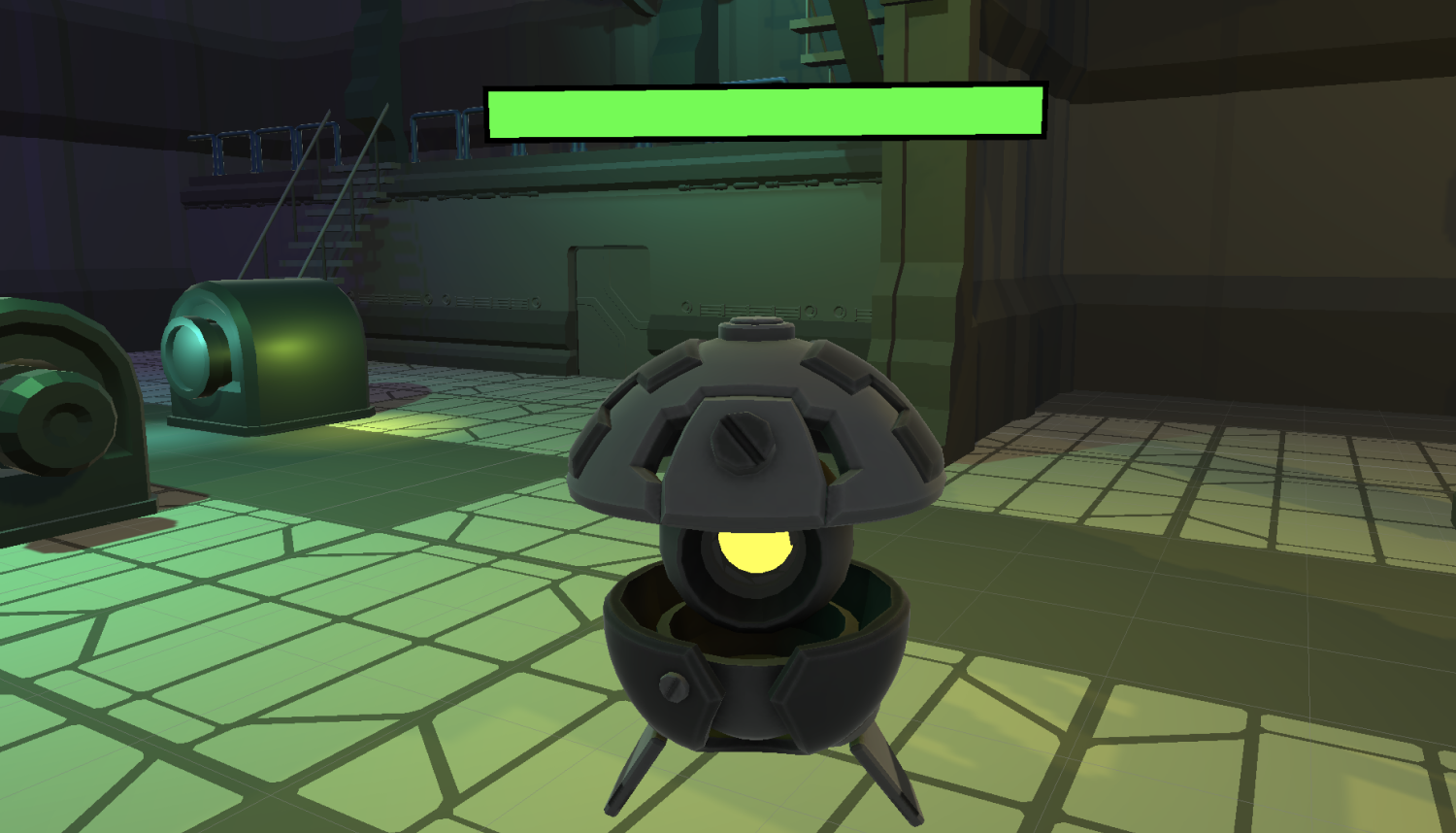}
  \hfill
  \includegraphics[width=.325\linewidth]{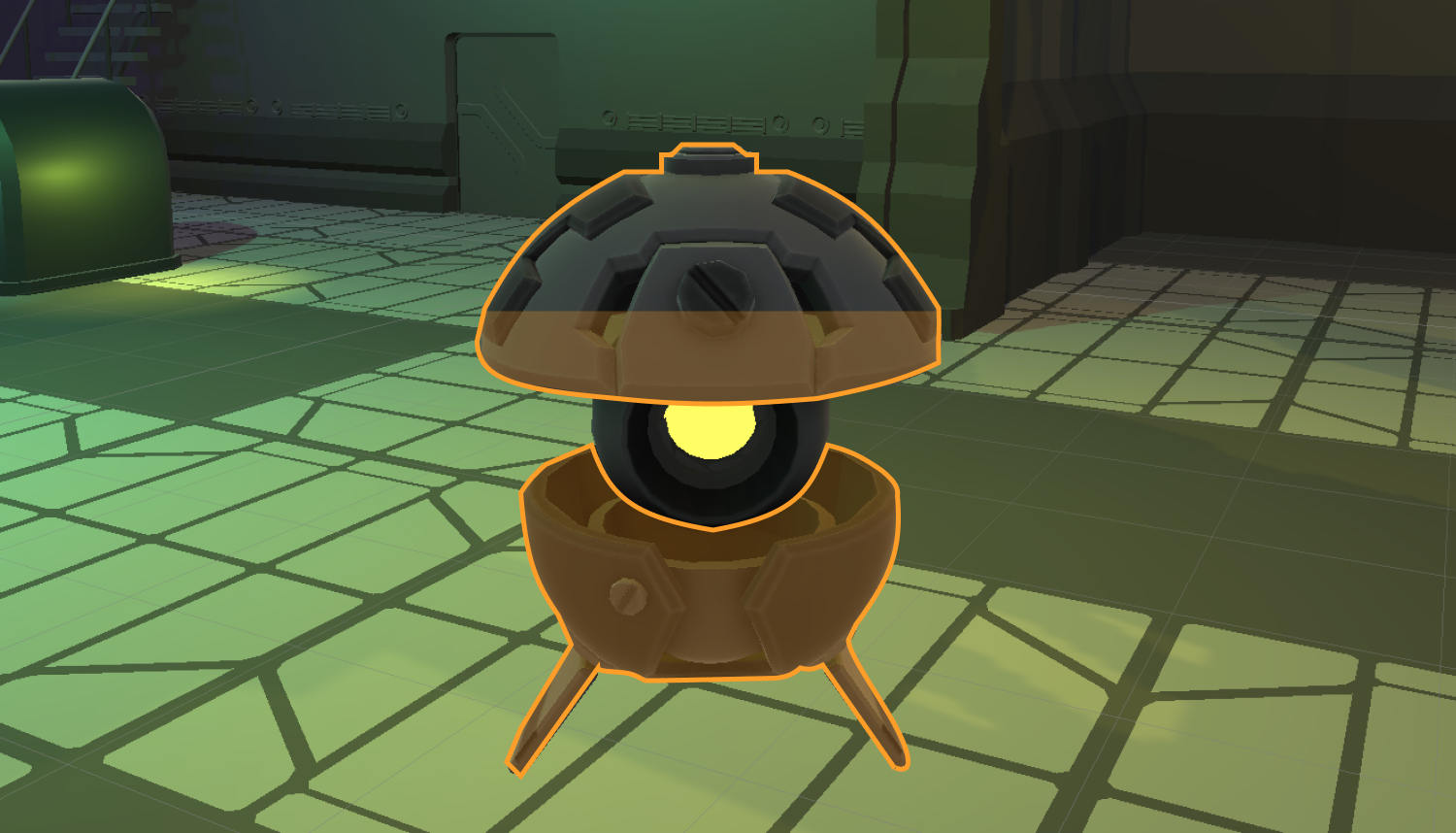}
\hfill
  \includegraphics[width=.325\linewidth]{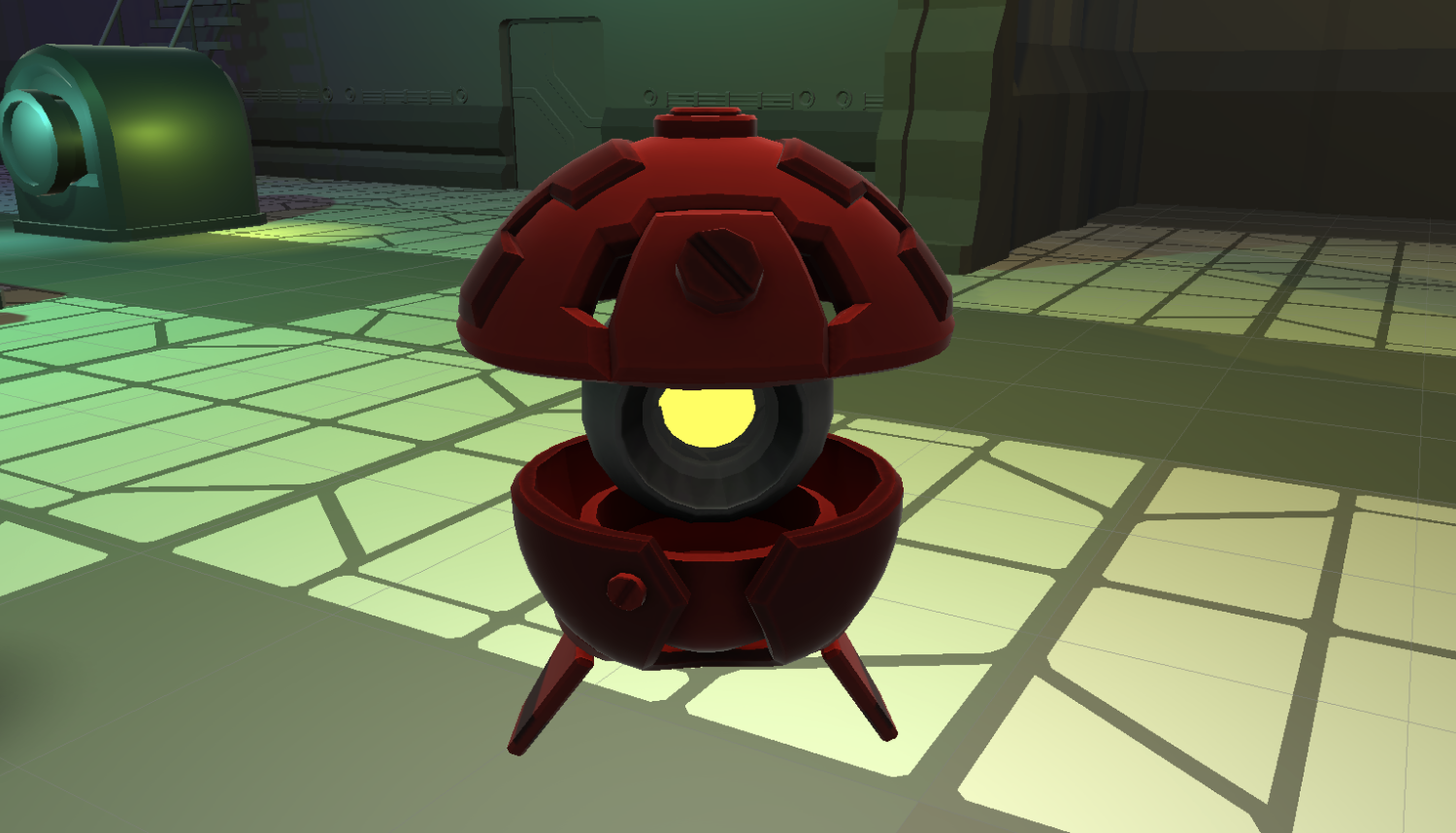}
  \caption{\label{fig:ex2}%
           Different embedding locations of visualizations in motion in a first person shooting game. Left: Health bar placed above the character's head. Middle: Health level expressed by color and silhouette overlapping the character. Right: Visualization integrated in the character's design (red color).
          }

\end{figure}

\vspace{-15pt}
\bpstart{Embedding locations:} refer to the spatial relationship of the visualization and the data referent. We found three different types of embedding locations (see \autoref{fig:ex3}: Right). 121/160 examples embedded visualizations \emph{around the data referent}. Visualizations were close to the referent for example above an object (see \autoref{fig:ex2}: Left), under the feet of a character, or above a checkpoint (a location). 26/160 visualizations showed full or partial \emph{overlap with the data referent} as shown for example in \autoref{fig:ex2}: Middle.
Finally, 13/160 visualizations were \emph{integrated with the data referent} permanently for example in the material color of the referent, see \autoref{fig:ex2}: Right. Although visualizations integrated or overlapped with the referent might look similar, integrated visualizations cannot be separated from the referent while overlays may be temporary. 

\vspace{-15pt}
\bpstart{Data dimensions:} indicate how many dimensions the visualization represented. Visualizations showing only a single information dimension were the most common (129/160), while only 31/160 samples showed multiple information dimensions (see \autoref{fig:ex3}: Left). Often additional dimensions were derived from a primary dimension. For example health bars often showed a single health percentage as a bar chart but also three different health states with a traffic light color scheme (healthy, medium, critical). 

%\begin{figure}[htb]
 % \centering
  % the following command controls the width of the embedded PS file
  % (relative to the width of the current column)
  %\includegraphics[width=\linewidth]{egPublStyle-EuroVis_full-short-stars-posters-2022/Fig 2.png}
  %
  %\caption{\label{fig:ex2}Example of multiple information dimension health bars.}
%\end{figure}

\begin{figure}[htb]
  \centering
     \includegraphics[width=.325\linewidth]{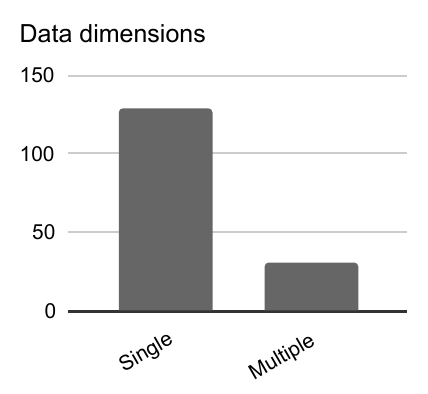}
  \includegraphics[width=.325\linewidth]{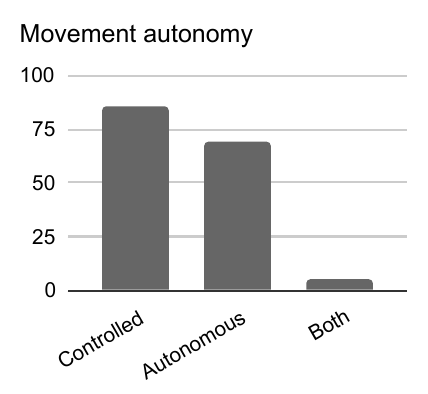}
  \caption{\label{fig:ex4}Counts for the number of examples including factors of the number of dimensions visualization and movement autonomy.}
\end{figure}

\vspace{-5pt}
\bpstart{Movement autonomy:} considers if the visualization was moving autonomously or was controlled by a player. For example, a  player can induce motion of a static enemy's health bar on the screen by moving his or her camera, for example via one's character. Autonomous movement of a visualization was not prevalent in video games, only 5/160 samples moved autonomously, while 86/160 were consistently controlled in some way only by the player and 69/160 depended on autonomous movement and the player's control (see \autoref{fig:ex4}: Right). 

\vspace{-2pt}

%-------------------------------------------------------------------------
\vspace{-20pt}
\section{Discussion and Future Work}
Our ultimate goal is to explore the impact of contextual factors in video games on visualizations in motion. We are currently in the first phase of this work. We surveyed visualizations in motions and analyzed them according to different dimensions. We are now looking for open-source games with the possibility to access their data in real-time. Furthermore, we plan to design our own visualizations in motion and then embed them into games. We may conduct empirical experiments to evaluate the readability of different visualizations in motion in the context of video games.

%-------------------------------------------------------------------------
\vspace{-10pt}
\section{Acknowledgement}
% uncomment for final version
This work was partly supported by the Agence Nationale de la Recherche (ANR), grant number ANR-19-CE33-0012.\\
\emph{Images credits: Screenshots came from the open source Unity FPS Microgame Template modified by the authors. Star icon by upklyak on freepik.com}

%-------------------------------------------------------------------------
% bibtex
\bibliographystyle{eg-alpha-doi}
\bibliography{Federica_poster}

% biblatex with biber
%\printbibliography            

%-------------------------------------------------------------------------
\end{document}